\documentclass[aps,prb,twocolumn,amsfonts,amssymb,showpacs]{revtex4}
\newcommand{\be}{\begin{equation}}
\newcommand{\bea}{\begin{eqnarray}}
\newcommand{\ba}{\begin{array}}
\newcommand{\ee}{\end{equation}}
\newcommand{\eea}{\end{eqnarray}}
\newcommand{\ea}{\end{array}}
\newcommand{\nn}{\nonumber}

\begin {document}

\title{Kondo Effects in Quantum Dots at Large Bias}
\author{Yu-Wen Lee$^1$ and Yu-Li Lee$^2$}
\affiliation{$^1$Department of Physics, National Taiwan University, Taipei,
    Taiwan, Republic of China\\
    $^2$Institute of Physics, Academia Sinica,  Taipei, Taiwan, Republic of China}
\date{Received 10 September 2001}
\begin{abstract}
Recently, the issue of whether the Kondo problem in quantum dots
at large bias is a weak-coupling problem or not has been raised.
In this paper, we revisit this problem by carefully analyzing a
corresponding model in the solvable limit --- the Emery-Kivelson
line, where various crossover energy scales can be easily
identified. We then try to extract the scaling behavior of this
problem from various physical correlation functions within the
spirit of ``poor man's scaling.'' Our conclusions support some
recent suggestions made by Coleman {\em et al.} [Phys. Rev. Lett.
{\bf 86}, 4088 (2001)], which are obtained by perturbative
analysis: The voltage acts as a cutoff of the renormalization
group flow for {\em only} half of the impurity so that the low-temperature
physics is controlled by a strong-coupling fixed
point. But the low-temperature response functions in general show
one-channel Kondo behaviors with two-channel Kondo
behaviors occurring only through proximity to a quantum critical
point.
\end{abstract}
\maketitle

\section{Introduction}

The Kondo problem is one of the best studied many-body problems in
condensed matter physics. Due to advances of nanotechnology, the
Kondo effect in quantum dot systems predicted by
theories\cite{theoriest} was observed in recent
experiments.\cite{exp} Although the quantum dot is intrinsically a
multilevel system, as the energy is much lower than the single-particle
level spacing, the system can be described by a single-impurity
Anderson model with the level spacing playing the role of
cutoff in this model. Moreover, in the Coulomb blockade regime
with an odd number of electrons, it can be mapped onto a two-lead
Kondo model with the following Hamiltonian: \cite{Glazman}
\begin{eqnarray}\label{hamtwo}
H & = & \sum_{\alpha {\bf k}\sigma} \varepsilon_{\alpha{\bf k}}
      c^\dag_{\alpha {\bf k} \sigma} c_{\alpha {\bf k} \sigma} +
      H_{\rm refl} + H_{\rm trans}, \nn\\
H_{\rm refl} & = & J_{R} \sum_{{\bf k},{\bf k}',\sigma,\sigma'}
      \left( c^\dag_{R{\bf k}\sigma} {\vec \sigma}_{\sigma \sigma'}
      c_{R {\bf k}' \sigma'} \right)\cdot {\vec S} \,\, + \,\,
      (R \to L), \nonumber \\
H_{\rm trans} & = & J_{LR} \sum_{{\bf k},{\bf k}', \sigma,\sigma'}
      \left(c^\dag_{L{\bf k}\sigma} {\vec \sigma}_{\sigma \sigma'}
      c_{R{\bf k}' \sigma'} \right)\cdot {\vec S} \,\, + \,\,
      (R \leftrightarrow L).
\end{eqnarray}
Here, $c^\dag_{\alpha {\bf k} \sigma}$ creates an electron in lead
$\alpha \in \left\{L,R\right\}$ with momentum ${\bf k}$ and spin
$\sigma$, and $J_L$, $J_R$, and $J_{LR}(=J_{RL})$ are positive
(antiferromagnetic) Kondo coupling constants between electrons and
the dot ($\vec{S}$). Energies $\varepsilon_{\alpha{\bf k}}=
\varepsilon_{{\bf k}}-e V_{\alpha} $, where $V_{\alpha}=\pm V/2$
are the potentials of the left- and right-hand leads. Besides,
derived from an Anderson model via a Schrieffer-Wolff
transformation\cite{Glazman}, the coupling constants obey
$|J_{LR}|^{2} = J_{L}J_{R}$.

Within the spirit of ``poor man's scaling,'' a recent paper by
Kaminski {\em et al.},~\cite{Glazman} shows that even in the
present nonequilibrium case, the low-temperature properties can
still be characterized by a {\em single} crossover energy scale
$T_K$ which is identified as the Kondo temperature of this
problem. Through extending the information gained from
perturbative renormalization group (RG) equations, Kaminski {\em
et al.}, use the ordinary one-channel Kondo fixed point to extract
low-temperature transport properties for $V\ll T_K$. However, for
$V\geq T_K$, the RG-improved perturbative calculation indicates
that the conductance will saturate at a value much smaller than
that in the unitary limit. This signals that the coupling between
different leads $J_{LR}$ stops growing for energy smaller than
$V$. One then wonders whether the above-mentioned one-channel Kondo (1CK) fixed point
can still be used to describe the low-temperature physics in the
latter case. Recently, to gain further insight into this problem,
Coleman {\em et al.},\cite{Coleman} have done a perturbative
calculation of the impurity magnetic susceptibility. The scaling
behaviors for various couplings they obtained are basically the
same as that obtained in Ref. \onlinecite{Glazman} in the high-temperature
region. However, for $T\ll V$, the flows of $J_{L,R}$
and $J_{LR}$ exhibit different behaviors: $J_{LR}$ stops growing
at the energy scale $V$ while $J_{L,R}$ continue to grow toward
strong coupling with a new Kondo temperature $T_K^*$. They
conjecture that the physics at $T<T_K^*$ is described by a 2CK
fixed point. (See also Ref. \onlinecite{XGWen}.)

In this paper, we shall revisit this problem by studying the model
in a solvable limit --- the generalized Emery-Kivelson
line.\cite{Hershfeld} This enables us to follow the RG flow all
the way from the perturbative regime down to the low-temperature
region, and the concept of universality guarantees that we can
obtain correct scaling at low temperature. Moreover, through
studying various correlation functions in details, we can see how
the impurity is screened (or unscreened) due to the presence of
finite bias. This approach is complementary to the perturbative
analysis in Ref. 4, which is supposed to be valid at high
temperature. Combined with the results obtained from the
perturbation theory, we reach a complete picture about the
behavior of the Hamiltonian (\ref{hamtwo}) at large bias: The
effect of the cotunneling term $H_{trans}$ is to generate a new
energy scale $\Gamma_{LR}$ even in the {\it channel-symmetric}
case ($J_L=J_R$). As $\Gamma_{LR}\ll T_K$, there exists a range of
temperature $\Gamma_{LR}\ll T\ll T_K$, where the uniform magnetic
susceptibility exhibits the 2CK behavior, i.e., the logarithmic
temperature dependence. On the contrary, it will show the 1CK
behavior as $\Gamma_{LR}\approx T_K$. This is the fundamental
difference between the present case and the ordinary 2CK fixed
point. Previous studies on ordinary 2CK problems\cite{emery}
revealed that the presence of an {\it unscreened} Majorana fermion
lies at the heart of the 2CK properties, which is the origin of the
logarithmic temperature dependence in various suscepetibilities.
In our case, the logarithmic temperature dependence is due to the
{\em partial} screening of a Majorana fermion. Although the low-temperature
behavior of the uniform susceptibility is similar to
that in the ordinary channel-asymmetric 2CK problem ($J_L\neq J_R$
and $H_{trans}=0$) and there are two characteristic energy scales
in both cases, the origins are distinct. In the ordinary
channel-asymmetric 2CK probelm, the new energy scale in addition
to the Kondo temperature is generated from the channel-asymmetry,
i.e., $J_L\neq J_R$, whereas in our case it arises from $H_{trans}$
and still exists even $J_L=J_R$.

The rest of the paper is organized as follows. We introduce the
solvable model in Sec. II. In Sec. III, the impurity Green
functions and the impurity contributions to the uniform magnetic
susceptibility are calculated. The last section is devoted to a
discussion and conclusions of our results.

\section{The solvable model}

After changing the notation $R(L)\to 1(2)$, we start with the
following Hamiltonian:\cite{lud,Fabrizio}
\begin{eqnarray*}
 { H} &=& { H}_0+{ H}_1+{ H}_2 \ ,
\end{eqnarray*}
where
\begin{eqnarray}
 { H}_0 &=& \sum_{\alpha , \sigma}\int dx :\psi^{\dagger}_{\alpha\sigma}
          (i\partial_x-V_{\alpha})\psi_{\alpha \sigma}: \ ,
          \nonumber \\
 { H}_1 &=& \hat{S}_z\sum_{\alpha}\lambda_{\alpha \parallel}
           \psi^{\dagger}_{\alpha}\frac{\sigma_3}{2}\psi_{\alpha}(0)
           \nn\\&+&\left(\hat{S}^{+}\sum_{\alpha}\frac{\lambda_{\alpha \perp}}
           {2}\psi^{\dagger}_{\alpha}\sigma_-\psi_{\alpha}(0)+{\rm H.c.}
           \right) \ , \nonumber  \\
 { H}_2 &=& \lambda_{LR\parallel}\hat{S}_z\psi^{\dagger}_2\frac{\sigma_3}
           {2}\psi_1(0)+\frac{\lambda_{LR\perp}}{2}\left[\hat{S}^{\dagger}
           \psi^{\dagger}_2\sigma_-\psi_1(0)\right.\nn\\&+&\left.\hat{S}^-\psi^+_2\sigma_+
           \psi_1(0)\right]+{\rm H.c.} \ .
\label{ham1}
\end{eqnarray}
Here $\alpha =1,2$, $\sigma = \uparrow ,\downarrow$, and $V_2=-V_1
=V/2$. For simplicity, we set $e=1$ and the Fermi velocity $v_F=1$
and $:\cdots:$ is the normal ordering with respect to the Fermi
surface in the absence of $V$. (Here we assume that Fermi
velocities on both leads are equal.)

The procedure to arrive at a solvable model is given in the
following. First we have to bosonize the Hamiltonian (\ref{ham1}).
To deal with Klein factors carefully, we employ bosonization
formulas on a finite length:\cite{delft}
\bea
 \psi_{\alpha \sigma}(x)&=& \frac{1}{\sqrt{2\pi a_0}} \ F_{\alpha \sigma}
          \exp{\{-i\Delta_L(\hat{N}_{\alpha \sigma}-P_0/2)x\}}
         \nn\\&&\times \exp{\{-i\sqrt{4\pi}\phi_{\alpha \sigma}\}} ,
\label{ferm1}
\eea
where $\hat{N}_{\alpha \sigma}$ is the number of $\psi_{\alpha
\sigma}$ fermions and $\Delta_L=2\pi/L$. Here $L$ is the system size
and $a_0$ is a short-distance cutoff. $P_0$ takes care of the
boundary conditions of fermion fields. $F_{\alpha \sigma}$'s are
Klein factors, which satisfy the commutation relations
$[F_{\alpha \sigma}, \hat{N}_{\alpha' \sigma'}]=\delta_{\alpha
\alpha'}\delta_{\sigma \sigma'}F_{\alpha \sigma}$, $\{F_{\alpha
\sigma}, F^{\dagger}_{\alpha'\sigma'}\}=2\delta_{\alpha\alpha'}\delta_{
\sigma \sigma'}$, and $\{F_{\alpha\sigma},F_{\alpha'\sigma'}\}
=0$. Next four bosonic fields are introduced\cite{emery} by
\begin{eqnarray}
 \phi_c &=& \frac{1}{2}(\phi_{1\uparrow}+\phi_{1\downarrow}+
         \phi_{2\uparrow}+\phi_{2\downarrow}) , \nonumber \\
 \phi_s &=& \frac{1}{2}(\phi_{1\uparrow}-\phi_{1\downarrow}+
         \phi_{2\uparrow}-\phi_{2\downarrow}) , \nonumber \\
 \phi_f &=& \frac{1}{2}(\phi_{1\uparrow}+\phi_{1\downarrow}-
         \phi_{2\uparrow}-\phi_{2\downarrow}) , \nonumber \\
 \phi_{sf} &=& \frac{1}{2}(\phi_{1\uparrow}-\phi_{1\downarrow}-
         \phi_{2\uparrow}+\phi_{2\downarrow}),
\label{ek1}
\end{eqnarray}
and the corresponding transformation on $\hat{N}_{\alpha \sigma}$
is
\begin{eqnarray}
 \hat{N}_c &=& \frac{1}{2}(\hat{N}_{1\uparrow}+\hat{N}_{1\downarrow}+
         \hat{N}_{2\uparrow}+\hat{N}_{2\downarrow}) , \nonumber \\
 \hat{N}_s &=& \frac{1}{2}(\hat{N}_{1\uparrow}-\hat{N}_{1\downarrow}+
         \hat{N}_{2\uparrow}-\hat{N}_{2\downarrow}) , \nonumber \\
 \hat{N}_f &=&\frac{1}{2}(\hat{N}_{1\uparrow}+\hat{N}_{1\downarrow}-
         \hat{N}_{2\uparrow}-\hat{N}_{2\downarrow}) , \nonumber \\
 \hat{N}_{sf} &=&\frac{1}{2}(\hat{N}_{1\uparrow}-\hat{N}_{1\downarrow}-
         \hat{N}_{2\uparrow}+\hat{N}_{2\downarrow}).
\label{ek2}
\end{eqnarray}
Here $\hat{N}_m\in {\cal Z}+P/2$ for $m =c,s,f,sf$ and $\sum_m
\hat{N}_m =0 \ {\rm mod} \ 2$. $P=0,1$ for the total number of
fermions being even and odd integers, respectively. After plugging
these into the Hamiltonian (\ref{ham1}), we perform the
Emery-Kivelson (EK) transformation:
$U=\exp{\{i\sqrt{4\pi}\hat{S}_z\phi_s(0)\}}$. To proceed, we also
introduce four more Klein factors $F_m$ with $m=c,s,f,sf$, which
satisfy the commutation relation: $[F_m,
\hat{N}_{m'}]=\delta_{mm'}F_m$. With the help of Eq. (\ref{ek2}),
the identification between $F_{\alpha \sigma}$ and $F_m$ can be
found. What we need is the following ones: $F^{\dagger}_{1\downarrow}F_{1
\uparrow}=F_sF_{sf}$, $F^{\dagger}_{2\downarrow}F_{2\uparrow}=F_s
F^{\dagger}_{sf}$, $F^{\dagger}_{2\downarrow}F_{1\uparrow}=F_fF_s$, and
$F^{\dagger}_{2\uparrow} F_{1\downarrow}=F_f F^{\dagger}_s$. Then, the
EK-transformed Hamiltonian is refermionized by the following
formulas:
\begin{eqnarray}
 d^{\dagger} &\equiv& F_s\hat{S}^+ , \ d \equiv F^{\dagger}_s\hat{S}^- , \nonumber \\
 \hat{S}_z &=& d^{\dagger} d-1/2 ,
\label{ferm2}
\end{eqnarray}
and
\begin{eqnarray}
 \Psi_m(x) &=& \frac{1}{\sqrt{2\pi a_0}} \ F_m\exp{\{-i\Delta_L
          (\hat{N}_m-1/2)x\}} \nn\\&&\times\exp{\{-i\sqrt{4\pi}\phi_m\}}, \
          m=f, sf , \nonumber \\
 \Psi_s(x) &=& \frac{1}{\sqrt{2\pi a_0}} \ F_s \ e^{i\pi d^{\dagger}d}\exp{\{
          -i\Delta_L(\hat{S}_T-1/2)x\}} \nn\\&&\times\exp{\{-i\sqrt{4\pi}\phi_s\}},
\label{ferm3}
\end{eqnarray}
with $\hat{S}_T=\hat{S}_z+\hat{N}_s$ being the total spin
operator. Finally, the structure of the Hamiltonian
can be further simplified by introducing Majorana fermions
\begin{eqnarray}
 \xi^1_m &=& \frac{\Psi_m+\Psi^{\dagger}_m}{\sqrt{2}}, \ \xi^2_m =
         \frac{\Psi_m-\Psi^{\dagger}_m}{\sqrt{2}i}, \ m=f,sf , \nonumber \\
 a &=& \frac{d+d^{\dagger}}{\sqrt{2}}, \ b=\frac{d-d^{\dagger}}{\sqrt{2}i}.
\label{ferm4}
\end{eqnarray}
Now, by taking $L \rightarrow \infty$ and ignoring
terms of $O(1/L)$ , we obtain
\begin{eqnarray*}
 { H}' &\equiv& U{ H}U^{\dagger}
          = {\bar{ H}}_0 +{ H}_{f}+{ H}_{sf}+{ H}_{int}+{\rm const},
\end{eqnarray*}
where
\begin{eqnarray}
 {\bar{ H}}_0 &=& \sum_{q>0}qb^{\dagger}_{qc}b_{qc}+\int dx :\Psi^{\dagger}_si\partial_x\Psi_s(x):,
          \nonumber \\
 { H}_f &=& \int dx \left(\sum_{\alpha=1,2}\frac{i}{2}
           \xi^{\alpha}_f\partial_x\xi^{\alpha}_f+iV\xi^1_f\xi^2_f
           \right) \nonumber \\
           & & -~i\sqrt{2\Gamma_{LR}} \ a\xi^2_f(0), \nonumber \\
 { H}_{sf} &=& \int dx \sum_{\alpha=1,2}\frac{i}{2}\xi^{\alpha}_{sf}
           \partial_x\xi^{\alpha}_{sf}-i\sqrt{2\Gamma_+} \ b
           \xi^1_{sf}(0) \nonumber \\
           & & +~i\sqrt{2\Gamma_-} \ a\xi^2_{sf}(0), \nonumber \\
 { H}_{int} &=& ab \ \left\{i\delta \lambda_1:\Psi^{\dagger}_s\Psi_s(0):-\delta
           \lambda_2\xi^1_{sf}\xi^2_{sf}(0) \right.\nonumber \\
           & &\left. +~\delta \lambda_3\xi^1_{sf}\xi^2_f(0)\right\}.
\label{ham2}
\end{eqnarray}
Here $b_{qc}=\sqrt{2q/L}\int dx e^{iqx}\phi_c(x)$ ($q>0$).
$\Gamma_{i} = \lambda^2_{i\perp}/(4\pi a_0)$ with $i=\pm , LR$.
 $\delta \lambda_1=\lambda_{+\parallel}-2\pi$ , $\delta \lambda_2=
\lambda_{-\parallel}$, and $\delta \lambda_3=\lambda_{LR
\parallel}$. $\lambda_{\pm i}=(\lambda_{1i}\pm \lambda_{2i})/2$
with $i=\perp , ||$. Notice that for symmetric dots, $\Gamma_-=0
=\delta \lambda_2$. The charge sector is completely decoupled from
the impurities and we will not consider it hereafter. From Eq.
(\ref{ham2}), we see that the EK line corresponds to $\delta
\lambda_{i}=0$ ($i=1,2,3$) and its Hamiltonian ${ H}_f+{ H}_{sf}$
can be solved exactly.

\section{Physical Observables}

Now we are in a position to compute impurity contributions to
the uniform magnetic susceptibility, which is defined by
$\chi_{imp}={\rm lim}_{L\to \infty}(\partial M/\partial
h|_{h=0}-L\chi_0)/L$ where $\chi_0$ is the Pauli susceptibility of
bulk electrons and $M=g\langle \hat{S}_T\rangle$ is the
magnetization. (We consider the case with gyromagnetic ratios on
the impurity site and bulk electrons being equal and denote it by
$g$.) From Eq. (\ref{ferm3}) and the operator product expansion (OPE)
of $\Psi^{\dagger}_s(z+ia)\Psi_s(z)$,
we have $\hat{S}_T=\int dx:\Psi^{\dagger}_s\Psi_s(x):+(P+1)/2$. The last
term does not depend on the magnetic field $h$ and we ignore it.

In the presence of external magnetic fields, there is an
additional term in the Hamiltonian, $\Delta { H}=-gh\hat{S}_T$,
which is invariant against the EK transformation. Performing the
EK transformation and refermionization successively, we obtain the
Hamiltonian describing noninteracting $\Psi_s$ fermions as
$ { H}_s = \int dx :\Psi^{\dagger}_s(i\partial_x-gh)\Psi_s(x):.$
The term linear in $h$ can be removed by the transformation
$\Psi_s(x)\to \Psi_s(x)e^{-ighx}$. As a result, ${ H}_s\to \int
dx :\Psi^{\dagger}_si\partial_x\Psi_s(x):+ {\rm const.}$. By noticing that
$:\Psi^{\dagger}_s\Psi_s(x): \to :\Psi^{\dagger}_s\Psi_s(x): +gh/(2\pi)$, the
magnetization becomes
\begin{equation}
 M = g\int dx \langle:\Psi^{\dagger}_s\Psi_s(x):\rangle +L\chi_0h,
\label{magn}
\end{equation}
while ${H}_{int}$ turns into
$ {\bar{H}}_{int} = {H}_{int}+i(gh/2\pi)\delta
             \lambda_1 ab .$~
It is straightforward to see that on the EK line the first term in
Eq. (\ref{magn}) vanishes. This leads to $\chi_{imp} =0$ and the
leading contribution to $\chi_{imp}$ {\it must} arise from
${\bar{H}}_{int}$.\cite{seng}

To calculate contributions to $\chi_{imp}$ given by ${\bar{
H}}_{int}$, we use the Keldysh formula.\cite{keld} Before plunging
into the calculation, we need impurity Green functions
$G_a(\omega)$ and $G_b(\omega)$ on the EK line:
\begin{eqnarray}
 G_a(\omega) &=& \left(\begin{array}{cc}
                 \frac{1}{\omega+i{\bar \Gamma}} &
                 -2i\frac{\Gamma_{LR}(n_{V-\omega}-n_{V+\omega})+
                 \Gamma_-(1-2n_{\omega})}{\omega^2+{\bar \Gamma}^2}
                 \\
                 0 & \frac{1}{\omega-i{\bar \Gamma}}
                 \end{array}\right) , \nonumber \\
 G_b(\omega) &=& \left(\begin{array}{cc}
                 \frac{1}{\omega+i\Gamma_+} &
                 -2i\frac{\Gamma_+}{\omega^2+\Gamma_+^2}(1-
                 2n_{\omega}) \\
                 0 & \frac{1}{\omega-i\Gamma_+}
                 \end{array}\right) ,
\label{gf2}
\end{eqnarray}
where ${\bar \Gamma}\equiv \Gamma_{LR}+\Gamma_-$ and $n_{\omega}$
is the Fermi distribution function. Here all Green functions are
defined in the Keldysh space as
\begin{eqnarray*}
G=\left(\begin{array}{cc}
 G_R & G_K \\
 0 & G_A
\end{array}\right)\, .\end{eqnarray*}
An important feature in Eq. (\ref{gf2}) is
that {\it in the presence of the cotunnelling term} ($\Gamma_{LR}\neq
0$) {\it $a$ fermions are always screened by electrons in leads at
low temperature ($\ll {\bar \Gamma}$) even for symmetric dots}.
However, we shall see later that the interpretation of ${\bar
\Gamma}$ depends on the ratio $|V|/\Gamma_+$. Finally, the bare
Green functions of $\Psi_s$ fermions are
\begin{equation}
 G^0_s(\omega ,p)=\left(\begin{array}{cc}
                  \frac{1}{\omega +p+i0^+} & -2\pi i(1-
                  2n_{\omega})\delta(\omega +p) \\
                  0 & \frac{1}{\omega +p-i0^+}
                  \end{array}\right).
\label{gf3}
\end{equation}

With the help of Eqs. (\ref{gf2}) and (\ref{gf3}) and ${\bar {
H}}_{int}$, we are able to find that in the perturbative expansion
of magnetization, the leading nonvanishing order is the one with
$\delta \lambda_1^2$. It consists of two terms: one independent of
$h$ while the other proportional to $h$. The former becomes zero
as $V=0$ and it does not contribute to the magnetic
susceptibility. Consequently, the latter gives the leading
behavior of $\chi_{imp}$ at low temperature. After performing
integrals over $p$ and $\omega$ and in terms of digamma function
$\psi (x)$, we have
\begin{eqnarray*}
 \chi_{imp}(T,V) &=& \frac{(g\delta \lambda_1)^2}{4\pi^3}
             \frac{1}{\Gamma_+-{\bar \Gamma}}(I_1+I_2),
\end{eqnarray*}
with
\begin{eqnarray}
 I_1 &=& \psi \left(\frac{1}{2}+\frac{\Gamma_+}{2\pi T}\right)-
         \psi \left(\frac{1}{2}+\frac{{\bar \Gamma}}{2\pi T}\right) ,
         \nonumber \\
 I_2 &=& \frac{\Gamma_{LR}}{\Gamma_++{\bar \Gamma}}{\rm Re}\left\{
         \psi \left(\frac{1}{2}+\frac{\Gamma_++iV}{2\pi T}\right)\right.\nn\\
         &&~~~~~~\left.-~
         \psi \left(\frac{1}{2}+\frac{{\bar \Gamma}+iV}{2\pi T}\right)
         \right\}.
\label{res1}
\end{eqnarray}
Here ${\rm Re}\{f(z)\}$ means the real part of $f(z)$. $I_1$ gives
the result of the ordinary 2CK problem and $\Gamma_+$
plays the role of Kondo temperature $T_K$ in that case without
channel asymmetry. Effects of finite bias on $\chi_{imp}$ are
through the function $I_2$ arising from scattering in the
spin-flip cotunneling channel.

Based on Eq. (\ref{res1}) and in terms of the asymptotic formula
of diagamma function $\psi(z)=\ln{z}-1/2z
-1/12z^2 +\cdots$, we can discuss the low temperature
behavior of $\chi_{imp}$. It depends on the ratio $R\equiv
\Gamma_+/{\bar
\Gamma}$. \\

(i) $R=O(1)$. When $T \ll \Gamma_+ , {\bar \Gamma}$, we have
\begin{eqnarray}
 I_1 &=& \ln{\left(\Gamma_+/{\bar\Gamma}\right)}+
     O\left([T/\Gamma_+({\bar \Gamma})]^2\right), \nonumber \\
 I_2 &\sim& \ln{f(V)}+O\left((T/\epsilon_1)^2\right),
\label{res4}
\end{eqnarray}
with $ f(V)=\sqrt{(\Gamma_+^2+V^2)/({\bar \Gamma}^2+V^2)}$.
Here the crossover energy $\epsilon_1=\Gamma_+({\bar \Gamma})$ and
$|V|$ for $|V|\ll \Gamma_+({\bar \Gamma})$ and $|V|\gg
\Gamma_+({\bar \Gamma})$, respectively. Thus, the leading behavior
of $\chi_{imp}$ is
\bea
 \chi_{imp}(T,V)&=&\frac{(g\delta \lambda_1)^2}{4\pi^3}\frac{1}
            {\Gamma_+-{\bar \Gamma}}\left[\ln{\left(\frac{\Gamma_+}
            {{\bar \Gamma}}\right)}\right.\nn\\&&\left.+
            \frac{\Gamma_{LR}}{\Gamma_++{\bar\Gamma}}\ln{f(V)}\right].
\label{res2}
\eea

(ii) $R\gg 1$. When ${\bar \Gamma}\ll T\ll
\Gamma_+$, we have
\begin{equation}
 \chi_{imp}(T,V)=\frac{(g\delta \lambda_1)^2}{4\pi^3}
            (A/\Gamma_+)\ln{(T^*_K/T)},
\label{res3}
\end{equation}
with $T^*_K\approx T_K\{1+[\Gamma_{LR}/2(\Gamma_++{\bar
\Gamma})]\ln{[1+(\Gamma_+/V)^2]}\}$ and $T_K$ for $|V|\gg T$ and
$|V|\ll T$, respectively. Here $A=1,1+\Gamma_{LR}/(\Gamma_++{\bar
\Gamma})$ for $|V|\gg T$ and $|V|\ll T$, respectively.
$T_K=c\Gamma_+$ with $c=2e^{\gamma}/\pi$ and $\gamma$ being the
Euler constant. Since $\Gamma_{LR}/\Gamma_+\ll 1$, $T^*_K$ is in
general with the same order as $T_K$. For extremely low
temperature $T\ll {\bar \Gamma}$, the behavior of $\chi_{imp}$
turns into that shown by Eq. (\ref{res2}). It is, however, a
crossover not a phase transition. Equations (\ref{res2}) and
(\ref{res3}) are the main results of this paper. We shall discuss
their implications now.

\section{Conclusions and discussions}

Combined with perturbative analysis in Ref. \onlinecite{Coleman},
we arrive at the following picture: In the case with $\Gamma_+ \approx
\bar\Gamma$, all couplings flow to the strong-coupling regime at
low energy as $|V|\ll \Gamma_+$ and both $a$ and $b$ fermions
are completely screened. It leads to 1CK behavior. This can
also be understood by noticing that the two-lead Hamiltonian
$H_f+H_{sf}$ in Eq. (\ref{ham2}) has the structure of {\em two}
copies of the 2CK problem where half of the impurity is screened
by $\psi_f$ and another half of the impurity is screened by
$\psi_{sf}$, and the voltage acts only on $\psi_f$. (For
simplicity, we take $\Gamma_ -=0$.) For $V\rightarrow 0$ and $
T\ll \Gamma_{LR}, \Gamma_+$, both channels will flow to strong
couplings and the system manages itself into a single-channel
Kondo problem, which can be easily seen by taking
$\Gamma_+=\Gamma_{LR}$ and rewriting the Hamiltonian through
defining a new fermion $\psi=(1/\sqrt 2)(\xi^2_f+i
\xi^1_{sf})$. On the other hand, for $|V|\gg \Gamma_+$, the flow
of $J_{LR}$ stops at the scale $V$ while $J_{L,R}$ still continue
to flow towards strong couplings without being affected by the
voltage. This is the origin of the scale $\epsilon_1$ appearing in
Eq. (\ref{res4}). It also leads to the fact that the Majorana
fermion $a$ cannot be completely screened and fluctuates with a
scale ${\bar \Gamma}$. Therefore, we expect that the 2CK behavior
emerges in some situation as shown in Eq. (\ref{res3}). This in
part explains the stability of this 2CK problem against the
perturbation $J_{LR}$. It is, however, {\em not} exactly
equivalent to an ordinary 2CK problem because the coupling
$J_{LR}$ below the scale $V$ is not irrelevant but {\em marginal}.
In other words, half of the impurity, the $a$ fermion, can be
viewed as a free fermion only at energy scales higher than ${\bar
\Gamma}$. As the temperature is far below it, the impurity behaves
as if it is completely screened.

The partially screened Majorana fermions also reveal themselves in
the conductance:\cite{Hershfeld}
\bea
  G(T,V)=G_U\frac{{\bar\Gamma}}{2\pi T} {\rm Re}
         \left\{\psi^{\prime}\left(\frac{1}{2} +\frac{\bar{\Gamma}+iV}
         {2\pi T}\right)\right\},
\eea
where $\psi^\prime(z)=d\psi/dz$ and $G_U=(\Gamma_{LR}/{\bar
\Gamma}){e^2}/{\pi}$ is the conductance in the unitary limit.
The low-temperature asymptotic behavior for $T\ll {\bar \Gamma}$
is \bea
 G(T,V) &=& G_U\frac{{\bar \Gamma}^2}{{\bar \Gamma}^2+V^2}
        +O\left((T/\epsilon_2)^2\right),
\label{current} \eea with the crossover energy $\epsilon_2=|V|$
and ${\bar \Gamma}$ for $|V|\gg {\bar \Gamma}$ and $|V|\ll {\bar
\Gamma}$, respectively. From Eq. (\ref{current}), we see that the
conductance exhibits very different behaviors for large and small
bias. For bias much smaller than the crossover energy
${\bar\Gamma}$, the conductance will reach the unitary limit, while
for $V\gg {\bar \Gamma}$, the voltage plays the role of a
crossover energy scale, and the conductance will saturate at a
much smaller value than that in the unitary limit. Within the
spirit of ``poor man's scaling," we can extract scaling behaviors
of the corresponding coupling: For $V\ll {\bar\Gamma}$, $J_{LR}$
will flow towards strong coupling and completely screen the
impurity. However, for $V\gg {\bar\Gamma}$, the RG flow will
terminate at the scale $V$ and the impurity can not be completely
screened. Note that the scaling behavior of $J_{LR}$
 gained here is completely consistent with
that obtained in Ref. \onlinecite{Coleman}. In addition, the
asymptotic behavior of Eq. (\ref{current}) is similar to that
appearing in Ref. \onlinecite{Glazman}, which reveals the
existence of some kind of universality in this nonequilibrium
problem.

Concerning the possible experimental realization of the above-mentioned
2CK behavior, we need to be close enough to that quantum
critical point. This requires a large ratio $R$ for the
renormalized scales, which is equivalent to the condition
$J_{LR}\ll J_{L,R}$. However, it is very difficult to achieve this
goal due to the constraint on the bare couplings, $J^2_{LR}\approx
J_L J_R$, and the logarithmic nature of the RG flow. Thus, unless
we can design a dot with a very small ratio for bare interlead
and intralead tunnelings, it seems that this window is too
narrow to observing the 2CK behavior.

Finally, we would like to mention that similar crossover
behaviors, i.e., the 2CK behavior at the intermediate-temperature
regime while the ordinary 1CK behavior in the extremely low-temperature
regime, have already appeared in the context of
channel-asymmetric 2CK problems.\cite{Fabrizio} Although some
technical details are not completely the same, the underlying
physical reasons are similar --- the appearance of a new energy
scale suppresses some scattering processes and changes the
direction of the RG flow. We would also like to point out that a
recent paper by Zvyagin~\cite{Zvyagin} discussed a similar problem
but in a totally different physical context by using the Bethe
ansatz. More precisely, Ref.~\onlinecite{Zvyagin} considers the
low-energy properties of conduction electrons hybridized with
localized $5f$ electrons. When the concentration of $5f$ electrons
is low, the magnetic susceptibility exhibits a similar crossover
behavior to our Eqs. (\ref{res2}) and (\ref{res3}). In that case,
the new energy scale in addition to the Kondo temperature arises
from the hybridization anisotropy. In a word, the most crucial
difference between the models considered in Refs.
\onlinecite{Fabrizio} and \onlinecite{Zvyagin} and the present one is
that the appearance of two crossover scales in the former case is
due to {\it channel anisotropy}, while in the latter case it is
due to {\it the cotunneling processes}. The mechanisms to generate
the new energy scale reveal their distinctions in the
channel-symmetric limit. In that limit, for the models considered
in Refs. \onlinecite{Fabrizio} and \onlinecite{Zvyagin}, the new
crossover scale {\em vanishes} and the low-temperature dynamics is
described by the {\it 2CK fixed point}. On the contrary, for the
Hamiltonian (\ref{hamtwo}) in the channel-symmetric limit,
the {\it two} crossover energy scales {\em never vanish}, and depending on the
relative magnitude of the two scales (a large difference induced
by large bias), the low-temperature dynamics is controlled either
by the 2CK or 1CK fixed points. However, it always exhibits
1CK-type behaviors when the temperature is far below both scales.

To sum up, we arrive at the following conclusions: (i) {\it
Regardless of the magnitude of bias} $V$, the low-temperature
physics of quantum dots coupled to two leads is controlled by {\it
a strong-coupling fixed point}. (ii) This strong-coupling fixed
point should exhibit the behavior of a {\it one-channel Kondo
fixed point} instead of a two-channel one. (iii) {\it There is a
nearby two-channel Kondo fixed point located at the unphysical
parameter space} ${\bar \Gamma}=0$ ( in the sense that
$\bar\Gamma=0$ requires $J_{LR}=0$). For dots with
${\bar\Gamma}\ll T_K$, {\it it controls the physics at the range
of temperature} ${\bar \Gamma}\ll T\ll T_K$.

\acknowledgments
 The work of Yu-Wen Lee is supported by the National
 Science Council of R.O.C. under Grant No. NSC89-2811-M002-0084.


\begin{thebibliography}{99}


\bibitem{theoriest} T. K. Ng and P. A. Lee, Phys. Rev. Lett. {\bf 61}, 1768 (1988)
 ; L. I. Glazman and M. E. Raikh, JETP Lett. {\bf 47}, 452 (1988).
 \bibitem{exp}D. Goldhaber-Gordon, {\it et al.}, Nature(London){\bf 391}, 156 (1998)
 ; S. M. Cronenwett, {\it et al.}, Science {\bf 281}, 540, (1998).
 \bibitem{Glazman} For a derivation of this Hamiltonian from the Anderson model through
  a {\em time-dependent} Schrieffer-Wolff transformation, see A. Kaminski, Yu. V. Nazarov,
  and L. I. Glazman, Phys. Rev. B {\bf 62}, 8154 (2000).
 \bibitem{Coleman} P. Coleman, C. Hooley, and O. Parcollet, Phys. Rev. Lett. {\bf 86},
             4088 (2001).
 \bibitem{XGWen} X. G. Wen, cond-mat/9812431 (unpublished).
 \bibitem{Hershfeld}A. Schiller and S. Hershfield, Phys. Rev. {\bf B58}, 14 978 (1998)
 \bibitem{emery} V. J. Emery and S. Kivelson, Phys. Rev. B {\bf 46},
                10 812 (1992).
 \bibitem{seng} A. M. Sengupta and A. Georges, Phys. Rev. B {\bf 49},
                10 020 (1994).
 \bibitem{lud} For mapping the Kondo Hamiltonian to a
               one-dimensional problem, see, for example, A. W. W.
               Ludwig, Int. J. Mod. Phys. B {\bf 8}, 347 (1994).
 \bibitem{Fabrizio} A similar Hamiltonian at zero-bias ( equilibrium ) and without the
                    cotunneling term has been studied by M. Fabrizio, A. O. Gogolin,
                    and Ph. Nozi\'eres, Phys. Rev. Lett. {\bf 74}, 4503 (1995) with
                    bosonization and N. Andrei and A. Jerez, {\it ibid} {\bf 74},
                    4507 (1995) with Bethe ansatz. However, the case they studied is
                    distinct from the present one which is intrinsincally a
                    {\em non-equlibrium} problem due to the finite bias.
 \bibitem{Zvyagin} A.A. Zvyagin, Phys. Rev B {\bf 63}, 014 503 (2000).
 \bibitem{delft} For details of bosonization on a finite length, see
                J. von Delft and H. Schoeller, Ann. Phys. (Leipzig) {\bf 7}, 225 (1998).
             (cond-mat/9805275.)
                Its applications to two-channel Kondo problems can
                be found in G. Zar\'and and J. von Delft, Phys. Rev.
                B {\bf 61}, 6918 (2000).
 \bibitem{keld} L.V. Keldysh, Zh. \'Eksp. Teor. Fiz. {\bf 47}, 1515
                (1964) [Sov. Phys. JETP {\bf 20}, 1018 (1965)].
                For discussions of the Keldysh formula, see, for example,
                A. M. Zagoskin, {\it Quantum Theory of Many-Body
                Systems} (Springer-Verlag New York, 1998) or
                M. Le Bellac, {\it Thermal Field Theory}
               (Cambridge University Press, Cambridge, England 1996).
\end{thebibliography}
\end{document}